\documentclass[prl,preprint,amsmath,amssymb]{revtex4}
\usepackage{graphicx}
\usepackage[justification=centering]{caption}

\begin{document}

\title{Quantifying quantum coherence and non-classical correlation based on Hellinger distance}
\author{Zhi-Xiang Jin$^{1}$}
\author{Shao-Ming Fei$^{1,2}$}

\affiliation{$^1$School of Mathematical Sciences, Capital Normal University,
Beijing 100048, China\\
$^2$Max-Planck-Institute for Mathematics in the Sciences, 04103 Leipzig, Germany}

\bigskip

\begin{abstract}
Quantum coherence and non-classical correlation are key features of quantum world. Quantifying coherence and non-classical correlation are two key tasks in quantum information theory.
First, we present a bona fide measure of quantum coherence by utilizing the Hellinger distance. This coherence measure is proven to
fulfill all the criteria of a well defined coherence measure, including the strong monotonicity in the resource theories of quantum coherence.
In terms of this coherence measure, the distribution of quantum coherence in multipartite systems is studied and a corresponding polygamy relation is proposed.
Its operational meanings and the relations between the generation of quantum correlations and the coherence are also investigated.
Moreover, we present Hellinger distance-based measure of non-classical correlation, which not only inherits the nice properties of the Hellinger distance including contractivity,
and but also shows a powerful analytic computability for a large class of quantum states. We show that there is an explicit
trade-off relation satisfied by the quantum coherence and this non-classical correlation.
\end{abstract}

\maketitle

\noindent{\bf Introduction}~
Quantum coherence and non-classical correlation are the key features of quantum world. Recent developments in our understanding of quantum coherence and non-classical correlation have come from the burgeoning field of quantum information science. One important pillar of the field is the study on quantification of coherence. Since the seminal work Ref. \cite{tmm} on defining
a good coherence measure in terms of the resource theory, quantum coherence has been extensively studied and applied to many quantum information processing Ref. \cite{tmm,spm,dg,ctmm,chs,jbdv,auhm,csb,ad,easm,em,eg,ir,irp,yxlc,ula,aer,mmtc,cmst,bukf}.

The relative entropy and $l_1$-norm coherence measures are two well-known measures of coherence, especially concerning the strong monotonicity property and the closed expressions.
In fact, different quantifications of coherence can greatly enrich our understanding of coherence. In particular, the distillable coherence Ref. \cite{wyd,yzcm}, the coherence of formation Ref. \cite{aj,wyd,yzcm}, the robustness of coherence Ref. \cite{nbcp}, the coherence measures based on entanglement Ref. \cite{auhm} and max-relative entropy of coherence Ref. \cite{bukf}, and the coherence concurrence Ref. \cite{qgy,dbq}, have been proposed and investigated.
For instance, the relative entropy coherence can be understood as the optimal rate for distilling a maximally coherent state from given states Ref. \cite{ad}. The robustness of coherence quantifies the advantage enabled by a quantum state in a phase discrimination task Ref. \cite{ctmm}. In addition,
the relations between coherence and path information Ref. \cite{bosp,bbch,bukf1}, the distribution of quantum coherence in multipartite systems Ref. \cite{rpjb}, the complementarity between coherence and mixedness Ref. \cite{ch,sbdp} have also been studied.

Besides the quantum entanglement, the quantification of other kind of quantum correlations like quantum discord \cite{lhvv,ohzw} has been also extensively investigated.
It has been shown that some quantum information processing tasks like assisted optimal state discrimination can be carried out without quantum entanglement,
if the one-side quantum discord is non-zero \cite{ekr,aac,bmma,ljm,tjkt,yjfs,adgv,ost}.
In this paper, we employ the quantum Hellinger distance to construct a new quantum coherence measure.
One prominent advantage of this measure is that it satisfies the strong monotonicity condition. Moreover, it has an analytic expression.
The relation between this coherence measure and the fidelity is derived, which gives rise further to the connection with the geometric measure of quantum coherence. We employ this coherence measure and present a clear polygamy relation that dominates the coherence distribution among multipartite systems.
Moreover, we present a measure of quantum correlation based on the Hellinger distance, which can be analytically solved for qubit-qudit states.
The tradeoff relation between the quantum coherence and quantum correlation is derived explicitly.

\noindent{\bf Coherence measure based on Hellinger distance}~
In the fixed computational basis $\{|i\rangle\}$ of a d-dimensional Hilbert space $H$,
the set of the incoherent states $\mathcal{I}$ has the form
$\delta=\sum_{i=1}^d\delta_i|i\rangle\langle i|$,
where $\delta_i\in[0,1]$ and $\sum_i\delta_i=1$. Baumgratz et al. Ref. \cite{tmm} proposed that any proper measure $C$ of coherence should satisfy the following conditions:
(C1) $C(\rho)\geq0$ for all quantum states $\rho$, and $C(\rho)=0$ if and only if $\rho\in \mathcal{I}$;
(C2a) Monotonicity under incoherent completely positive and trace preserving maps (ICPTP) $\Psi$, i.e., $C(\rho) \geq C(\Psi(\rho))$;
(C2b) Monotonicity for average coherence under subselection based on measurements outcomes: $C(\rho)\geq\sum_i p_iC(\rho_i )$, where $\rho_i= K_i\rho K_i^\dagger/p_i$ and $p_i=\mathrm{Tr}( K_i\rho K_i^\dagger)$ for all ${K_i}$ with $\sum_iK_i^\dagger K_i=I$ and $K_i\mathcal{I}K_i^\dagger\subseteq \mathcal{I}$;
(C3) Non-increasing under mixing of quantum states (convexity), i.e., $\sum_ip_iC(\rho_i)\geq C(\sum_ip_i\rho_i)$, for any ensemble $\{p_i, \rho_i\}$.
Note that conditions (C2b) and (C3) automatically imply condition (C2a). The condition (C2b) is important as it allows for sub-selection based on measurement outcomes, a process available in well controlled quantum experiments Ref. \cite{tmm}. It has been shown that the relative entropy measure and the $l_1$-norm measure of coherence satisfy all these conditions. However, the measure of coherence induced by the squared Hilbert-Schmidt norm satisfies conditions (C1), (C2a) and (C3), but not (C2b). Recently, the fidelity measure of coherence is proved to be a measure of coherence which does not satisfy the condition (C2b) Ref. \cite{FHS}.

In the following we first introduce a Hellinger distance based measure of coherence and show that it is a bona fide measure of quantum coherence.
Let $D_H(\rho,\delta)$ denote the Hellinger distance between two states $\rho$ and $\delta$, $D_H(\rho,\delta)=\mathrm{Tr}(\sqrt{\rho}-\sqrt{\delta})^2$.

{\bf [Theorem 1]}. The quantum coherence $C_H(\rho)$ of a state $\rho$ quantified by
\begin{eqnarray}\label{def1}
C_H(\rho)={\min}_{\delta\in \mathcal{I}}D_H(\rho,\delta)
\end{eqnarray}
is a well-defined measure of coherence.

{\bf [Proof]}. Set $f(\rho, \sigma)=\mathrm{Tr}\sqrt{\rho}\sqrt{\sigma}$. Then $C_H(\rho)$ can be expressed as
$C_H(\rho)=2\left(1-\mathrm{max}_{\delta\in \mathcal{I}}f(\rho,\delta)\right)$.
As an incoherent state $\delta$ can be explicitly written as
$\delta=\sum_{k=0}^{d-1}\delta_{k}|k\rangle\langle k|$, we have
\begin{eqnarray}\label{pfle2}
f(\rho,\delta)=\sum_{k=0}^{d-1}\langle k|\sqrt{\rho}|k\rangle\sqrt{\delta_{k}}
=M \sum_{k=0}^{d-1}\frac{\langle k|\sqrt{\rho}|k\rangle}{M}\sqrt{\delta_{k}},
\end{eqnarray}
with $M=\sqrt{\sum_{k=0}^{d-1}\langle k|\sqrt{\rho}|k\rangle^2}$. According to the Cauchy-Schwarz inequality, we have
\begin{eqnarray}\label{pfle3}
\left(\sum_{k=0}^{d-1}\frac{\langle k|\sqrt{\rho}|k\rangle}{M}\sqrt{\delta_{k}}\right)^2\leq \left(\sum_{k=0}^{d-1}\frac{\langle k|\sqrt{\rho}|k\rangle^2}{M^2}\right)\left(\sum_{k=0}^{d-1}\delta_{k}\right)=1,
\end{eqnarray}
with the inequality saturated when
\begin{eqnarray}\label{pfle4}
\sqrt{\delta_{k}}=\frac{\langle k|\sqrt{\rho}|k\rangle}{M}.
\end{eqnarray}
Substituting Eq. (\ref{pfle3}) into Eq. (\ref{pfle2}), one gets
$f(\rho,\delta)\leq M$. Therefore,
\begin{eqnarray*}\label{pfle5}
\mathrm{max}_{\delta\in \mathcal{I}}f(\rho,\delta)= M=\sqrt{\sum_{k=0}^{d-1}\langle k|\sqrt{\rho}|k\rangle^2}.
\end{eqnarray*}
From Eq. (\ref{pfle4})
$\delta=\delta_0=\sum_k\frac{\langle k|\sqrt{\rho}|k\rangle^2}{\sum_{k'}\langle k'|\sqrt{\rho}|k'\rangle^2}|k\rangle\langle k|$ is the optimal incoherent state that attains the maximal value of $f(\rho,\delta)$.

It is easy to find that $\mathrm{min}_{\delta\in \mathcal{I}}\mathrm{Tr}(\sqrt{\rho}-\sqrt{\delta})^2=0$, iff $\rho$ is an incoherent state. As $f(\rho,\sigma)$ is concave Ref. \cite{LSL}, $C_H(\rho)$ is convex under mixing states. That is, the criteria (C1) and (C3) are automatically satisfied. In addition,
since the coherence measure is convex, the monotonicity on selective incoherent completely positive and trace preserving mapping (ICPTP) (strong monotonicity) automatically implies the monotonicity on ICPTP.

Next we prove that $C_H(\rho)$ satisfies (C2a) -- the strong monotonicity.
Let $\delta^0$ denote the optimal incoherent state achieving the maximal value of $f(\rho,\delta)$.
Let $\Lambda= \{O_i\}$ be the incoherent selective quantum operations given by Kraus operators
$\{O_i\}$, with $\sum_{i=0}O_i^{\dagger}O_i=I$.
Under the operation $\Lambda$ on a state $\rho$, the post-measurement ensemble is given by $\{p_i, \rho_i\}$ with $p_i = \mathrm{Tr}O_i\rho O_i^{\dagger}$ and $\rho_i=O_i\rho O_i^{\dagger}/p_i$. Therefore, the average coherence is given given by
\begin{eqnarray}\label{pfth11}
\sum_{i=1}p_iC_H(\rho_i)=2\left(1-\sum_ip_i[\mathrm{max}_{\delta_i\in \mathcal{I}}f(\rho_i,\delta_i)]\right).
\end{eqnarray}
Since the incoherent operation cannot generate coherence from an incoherent state, for the optimal incoherent state $\delta^0$, we have $\delta_i^0=O_i\delta^0 O_i^{\dagger}/q_i\in \mathcal{I}$ with $q_i=\mathrm{Tr}O_i\delta^0 O_i^{\dagger}$ for any incoherent operation $O_i$. Thus for such a particular $\delta_i^0$, one has
$f(\rho_i,\delta_i^0)\leq \mathrm{max}_{\delta_i\in \mathcal{I}}f(\rho_i,\delta_i)$.
Hence Eq. (\ref{pfth11}) can be rewritten as
\begin{eqnarray}\label{c1}
\sum_{i=1}p_iC_H(\rho_i)\leq 2\left(1-\sum_ip_if(\rho_i,\delta_i^0)\right).
\end{eqnarray}

Let $\{a_k\}, \{b_k\}$ are two real numbers,
$a_1\leq a_2\leq \cdots \leq a_n$, $b_1\leq b_2\leq \cdots \leq b_n$.
By using the fact that under
arbitrary permutations $\sigma$: $[n] \to [n]$ Ref. [36],
\begin{eqnarray*}\label{}
\sum_{k=1}^na_kb_{n-k+1}\leq \sum_{k=1}^na_kb_{\sigma(k)}\leq \sum_{k=1}^na_kb_k,
\end{eqnarray*}
We have
\begin{eqnarray}\label{c2}
\sum_ip_if(\rho_i,\delta_i^0)\geq \sum_i\sqrt{p_iq_i}f(\rho_i,\delta_i^0),
\end{eqnarray}
where $p_i=\mathrm{Tr}O_i\rho O_i^{\dagger}$, $q_i=\mathrm{Tr}O_i\delta^0 O_i^{\dagger}$, and $\rho_i={O_i\rho O_i^{\dagger}}/{p_i}$, $\delta^0 _i={O_i\delta^0  O_i^{\dagger}}/{q_i}$.
Therefore,
\begin{eqnarray*}\label{c3}
\sum_{i=1}p_iC_H(\rho_i)&&\leq 2\left(1-\sum_ip_if(\rho_i,\delta_i^0)\right)\nonumber\\
&&\leq 2\left(1-\sum_i\sqrt{p_iq_i}f(\rho_i,\delta_i^0)\right)\nonumber\\
&&\leq 2\left(1-f(\rho,\delta^0)\right)=C_H(\rho),
\end{eqnarray*}
where the first inequality is due to Eq. (\ref{c1}). From Eq. (\ref{c2}), we can get the second inequality. The third inequality is due to $f(\rho, \sigma)\leq \sum_i\sqrt{p_iq_i}f(\rho_i,\sigma_i)$ [37].
It shows the strong monotonicity. The monotonicity is directly given by the
convexity of $C_H(\rho)$, $C_H(\rho)\geq C_H(\sum_{i=1}p_i\rho_i)=C_H(\Lambda(\rho))$.
\quad $\Box$

{\it Remark.} From the proof of Theorem 1, one sees that
$C_H(\rho)={\min}_{\delta\in \mathcal{I}}D_H(\rho,\delta)=2\left(1-\sqrt{\sum_k\langle k|\sqrt{\rho}|k\rangle^2}\right)$ under fixed computational basis.
There is a quantitative relation between the coherence measure $C_H$ based on Hellinger distance and the coherence measure $C$ introduced in Ref. \cite{YCS}: $C_H=2(1-\sqrt{1-C})$.
However, that $C$ is a coherence measure does not imply that $C_H$ is a coherence measure too. Generally a function of $C$ does not satisfy
all the necessary conditions of a proper coherence measure, even if $C$ is a well defined coherence measure.
Therefore, proving $C_H$ to be a bona fide measure of coherence is necessary. In addition, different measures give rise to different operational meanings
and different physical implications. As will be seen, $C_H$ is connected with fidelity, while $C$ in Ref. \cite{YCS} is related to quantum metrology.

\noindent{\bf Relation between coherence $C_H$ and fidelity}~
The geometric measure of coherence $C_g$ Ref. \cite{ast} is defined by
\begin{eqnarray*}\label{cg}
C_g(\rho)=1-\mathrm{max}_{\delta\in \mathcal{I}}F(\rho,\delta),
\end{eqnarray*}
where $F(\rho,\delta)=\mathrm{Tr}\sqrt{\rho^{1/2}\delta\rho^{1/2}}$ is the fidelity of two density operators $\rho$ and $\delta$.

It is direct to show that for arbitrary states $\rho$ and $\delta$,
\begin{eqnarray}\label{f1}
f(\rho,\delta)=\mathrm{Tr}(\sqrt{\rho}\sqrt{\delta})\leq \mathrm{max}_U\mathrm{Tr}(\sqrt{\rho}\sqrt{\delta}U),
\end{eqnarray}
and
\begin{eqnarray}\label{f2}\nonumber
\mathrm{Tr}(\sqrt{\rho}\sqrt{\delta}U)&&=\mathrm{Tr}(|\sqrt{\rho}\sqrt{\delta}|VU)\\\nonumber
&&=\mathrm{Tr}(|\sqrt{\rho}\sqrt{\delta}|^{1/2}|\sqrt{\rho}\sqrt{\delta}|^{1/2}VU)\\\nonumber
&&\leq \sqrt{\mathrm{Tr}(|\sqrt{\rho}\sqrt{\delta}|\mathrm{Tr}(U^\dagger V^\dagger|\sqrt{\rho}\sqrt{\delta}|VU)}\\\nonumber
&&=\mathrm{Tr}|\sqrt{\rho}\sqrt{\delta}|=\mathrm{Tr}\sqrt{\rho^{1/2}\delta\rho^{1/2}}\\
&&=F(\rho,\delta),
\end{eqnarray}
where $A=|A|V$ is the polar decomposition of $A$. The inequality is due to Cauchy-Schwarz inequality.  The equality can be attained by choosing $U=V^\dagger$.

From Eq. (\ref{f1}) and Eq. (\ref{f2}), we obtain that $F(\rho,\delta)\geq f(\rho,\delta)$.
Therefore, by the definition of geometric measure of coherence $C_g$, we get $C_H(\rho)\geq C_g(\rho)$.
Moreover, $C_H(\rho)$ is lower bounded by the minimal fidelity of $\rho$ and incoherent states.

In addition, from the above derivation in Eq. (\ref{f2}), we have
$F(\rho,\delta)=\mathrm{max}_U\mathrm{Tr}(\sqrt{\rho}\sqrt{\delta}U)$.
In particular, for any pure state $|\psi\rangle$, we get
\begin{eqnarray}\label{f3}\nonumber
F(|\psi\rangle\langle\psi|,\delta)&&= \mathrm{max}_U \langle\psi|\sqrt{\delta}U|\psi\rangle)=\mathrm{max}_U \langle\psi|\sqrt{\delta}|\psi\rangle \langle\psi|U|\psi\rangle)\\\nonumber
&&=\langle\psi|\sqrt{\delta}|\psi\rangle=f(|\psi\rangle\langle\psi|,\delta).
\end{eqnarray}
Therefore, $C_H(|\psi\rangle)=C_g(|\psi\rangle)$, which is given by the minimal fidelity of $|\psi\rangle$ and the incoherent states.

As an example, let us consider the maximally coherent mixed states Ref. \cite{sbdp},
\begin{eqnarray}\label{mcms}\nonumber
\rho_m=p|\psi\rangle_4\langle\psi|+\frac{(1-p)}{4}{I}_4,
\end{eqnarray}
where $0<p\leq 1$, ${I}_4$ is the $4\times 4$ identity matrix, and $|\psi\rangle_4=\frac{1}{2}\sum_{i=1}^4|i\rangle$ is the maximally coherent state.
We have $C_H(\rho_m)=2(1-\frac{\sqrt{2+2p}}{4})$. By Ref. \cite{zhj}, one obtains $C_g(\rho_m)=1-\frac{3}{4}\sqrt{1-p}-\frac{1}{4}\sqrt{1+3p}$, with $\mathrm{max}_{\delta_i\in \mathcal{I}}F(\rho_m, \delta)=\frac{3}{4}\sqrt{1-p}+\frac{1}{4}\sqrt{1+3p}$.
We can see that $C_H(\rho_m)\geq C_g(\rho_m)$, and $C_H(\rho_m)=C_g(\rho_m)$ when $p=1$, i.e. $\rho_m$ is the maximally coherent state. We can also see that $C_H(\rho_m)$ is a lower bound of the maximal fidelity $\mathrm{max}_{\delta_i\in \mathcal{I}}F(\rho_m, \delta)$, see Fig. 1

\begin{figure}
  \centering
  \includegraphics[width=10cm]{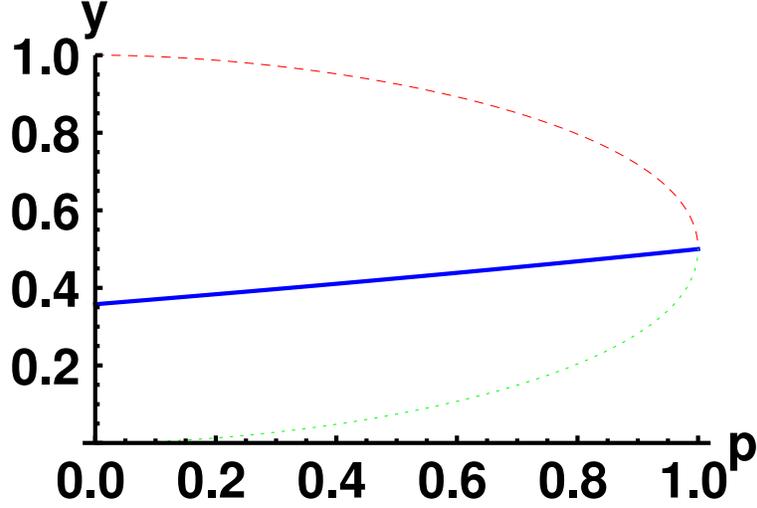}\\
  \caption{Solid (blue) line for $y=C_H(\rho_m)/2$; dashed (red) line for $y=\mathrm{max}_{\delta_i\in \mathcal{I}}F(\rho_m, \delta)$; dotted (green) line for $y=C_g(\rho_m)$, respectively.}\label{2}
\end{figure}

\noindent{\bf Distribution of coherence in multipartite systems}~
In the following, we consider the distribution of coherence among multipartite systems in terms of the coherence measure $C_H(\rho)$.
This essentially requires to extend the coherence to multipartite systems and establish the trade-off relation between the coherence among different subsystems. Using the similar methods of Ref. \cite{YCS}, we have the following results.

{\bf [Proposition 1]}. For a bipartite pure state $|\psi\rangle_{AB}$, we have
\begin{eqnarray}\label{th2}
1-\frac{1}{2}C_H(|\psi\rangle_{AB})\leq [1-\frac{1}{2}C_H(\rho_A)][1-\frac{1}{2}C_H(\rho_B)],
\end{eqnarray}
where $\rho_{A}$ ($\rho_{B}$) denote the reduced density matrix for partite $A$ ($B$).
The inequality Eq. (\ref{th2}) is saturated for product states.

{\bf [Proposition 2]}. For bipartite mixed states $\rho_{AB}$ with reduced density matrices $\rho_A$ and $\rho_B$, the coherences satisfy the following relations,
\begin{eqnarray}
[1-\frac{1}{2}C_H(\rho_A)]^2[1-\frac{1}{2}C_H(\rho_B)]^2\geq(\mathrm{Tr}\rho_{AB}^2-C_{l_2}(\rho_{AB}))\geq \lambda_{\mathrm{min}}\,[1-\frac{1}{2}C_H(\rho_{AB})]^2,\label{co11}\\
\label{copf114}
[1-\frac{1}{2}C_H(\rho_{A})][1-\frac{1}{2}C_H(\rho_{B})]\geq \frac{1}{{c_s}} [1-\frac{1}{2}C_H(\rho_{AB})]^2\nonumber,
\end{eqnarray}
where $\lambda_{\mathrm{min}}$ denotes the minimal nonzero eigenvalue of $\rho_{AB}$, $C_{l_2}(\rho)=\sum_{i\ne j}|\rho_{ij}|^2$ is the $l_2$-norm coherence measure of $\rho$, $c_s=\sqrt{\sum_i^r[1-\frac{1}{2}C_H(\rho_{A_i})]^2\sum_i^r[1-\frac{1}{2}C_H(\rho_{B_i})]^2}$, $r$ is the rank of $\rho_{AB}$ and $\rho_{A_i}$, $\rho_{B_i}$ denote the reduced density
matrices of the $i$-th eigenstate of $\rho_{AB}$.

In Ref. \cite{cmst}, it has been shown that the tradeoff relation satisfied by the coherence among different subsystems depends on states. Namely,
the coherence satisfies monogamous relations for some states, but polygamous ones for other states. Proposition 2 gives a general way
in describing certain properties of polygamy that satisfied by any states.
From Proposition 2, it can be found that the coherence of a subsystem is not limited by the coherence of the composite system.
A trivial case is that an incoherent composite pure state implies vanishing coherence of the subsystems.
However, a composite state with large coherence does not restrict the coherence of the subsystems: the subsystems may also have large coherence,
which is different from the monogamy of quantum entanglement. For the maximally coherent state, $|\psi\rangle_{AB}=\frac{1}{3}\sum_{i,j=0}^2|ij\rangle$,
we have $C_H(|\psi\rangle_{AB})=\frac{4}{3}$, $C_H(\rho_A)=C_H(\rho_B)=2(1-\sqrt{\frac{1}{3}})$, which is the maximal coherence in three-dimensional space corresponding to the reduced states $\rho_A=\rho_B=\frac{1}{9}\sum_{i,j=0}^2|i\rangle\langle j|$. This example implies that a subsystem with relatively large coherence does not restrict its ability to
interact with another subsystem and gives rise to large coherence of the whole composite system. These are the manifestations of the so-called polygamy.

One can easily find that Eq. (\ref{co11}) reduces to the relation Eq. (\ref{th2}) if $\rho_{AB}$ is a pure state
($\lambda_{\mathrm{min}}=1$).
Now we consider $N$-partite quantum states $\rho_{AB\cdots N}$.
Let $\alpha$ be a subset of $S=\{A,B,\cdots,N\}$ and $\rho_{\alpha}$ the corresponding reduced density matrix.

{\bf [Proposition 3]}. Let $\alpha_i \subset \{A,B,\cdots,N\}$ such that
$\alpha_i\cap \alpha_j=\delta_{ij}$ and $\sum_{i=1}\alpha_i=S$.
For an $N$-partite quantum state $\rho_{AB\cdots N}$, the coherences satisfy
\begin{eqnarray*}\label{co2a}
\prod_i[1-\frac{1}{2}C_H(\rho_{\alpha_i})]\geq \sqrt{\lambda_M}[1-\frac{1}{2}C_H(\rho_{AB\cdots N})],
\end{eqnarray*}
\begin{eqnarray*}\label{co2b}
\prod_i[1-\frac{1}{2}C_H(\rho_{\alpha_i})]^{n_i}\geq \frac{1}{c_{sT}}[1-\frac{1}{2}C_H(\rho_{AB\cdots N})]^2,
\end{eqnarray*}
where $n_i$, $\lambda_M$ and $c_{sT}$ are determined by
Proposition 2, depending on the detailed partitions $\alpha_i$.

\noindent{\bf Quantum correlations based on Hellinger distance}~
In Ref. \cite{46,47} it has been shown that any degree of coherence in some reference basis can be converted to entanglement
via incoherent operations. And in Ref. \cite{248} a general relation between coherence and entanglement under any measures
has been established. Here we show that coherence can be also converted to non-classical correlations via incoherent operations.

We first introduce the non-classical correlation $D(\rho_{AB})$ based on the Hellinger distance for bipartite states $\rho_{AB}$,
\begin{eqnarray}\label{defone}
D(\rho_{AB})=\mathrm{min}_{\{\Pi_A^k\}}\sum_kD_H(\rho_{AB},(\Pi_A^k\otimes I_B)\rho_{AB}(\Pi_A^k\otimes I_B)),
\end{eqnarray}
where ${I}_B$ is the identity operator on system $B$,
$D_H(\rho_{AB},(\Pi_A^k\otimes I_B)\rho_{AB}(\Pi_A^k\otimes I_B))=\mathrm{Tr}\left(\sqrt{\rho_{AB}}-\sqrt{(\Pi_A^k\otimes I_B)\rho_{AB}(\Pi_A^k\otimes I_B)}\right)^2$,
$\Pi_A^k=|k\rangle_A\langle k|$ denotes the projective measurement on subsystem $A$.

We first show some properties of $D(\rho_{AB})$ and demonstrate that it is a well defined measure of quantum correlation.

1) $D(\rho_{AB})$ is invariant under local unitary operations. We have
\begin{eqnarray*}\label{}
&&D[(U_A\otimes U_B)\rho_{AB}(U_A\otimes U_B)^\dagger]\nonumber\\
&&=\min_{\{\Pi_A^k\}}\sum_kD_H[(U_A\otimes U_B)\rho_{AB}(U_A\otimes U_B)^\dagger,  (\Pi_A^k\otimes I_B)(U_A\otimes U_B)\rho_{AB}(U_A\otimes U_B)^\dagger(\Pi_A^k\otimes I_B)]\nonumber\\
&&=\min_{\{\Pi_A^k\}}\sum_kD_H[\rho_{AB},  (U_A\otimes U_B)^\dagger(\Pi_A^k\otimes I_B)(U_A\otimes U_B)\rho_{AB}(U_A\otimes U_B)^\dagger(\Pi_A^k\otimes I_B)(U_A\otimes U_B)]\nonumber\\
&&=\min_{\{\Pi_A^k\}}\sum_kD_H[\rho_{AB},  (U_A^\dagger\Pi_A^k U_A\otimes I_B)\rho_{AB}(U_A^\dagger\Pi_A^k U_A\otimes I_B)]\nonumber\\
&&=D(\rho_{AB}),
\end{eqnarray*}
as minimizing over the local measurements $\{\Pi_A^k\}$ is obviously equivalent to do it over the ones rotated by $U_A$.

2) $D(\rho_{AB})$ is contractive under completely positive and
trace-preserving maps $\Psi_B$ on $B$. Note that $D_H(\rho_{AB})$ is contractive under completely positive and trace-preserving maps $\Psi_B$ on $B$ \cite{LSL}, $D_H(\rho_{AB}, \delta_{AB})\geq D_H((I_A\otimes \Psi_B)\rho_{AB}, \delta_{AB}) $. Let $\{\tilde{\Pi}_A^k\}$ be the optimal measurement such that $D(\rho_{AB})=\sum_kD_H(\rho_{AB},(\tilde{\Pi}_A^k\otimes I_B)\rho_{AB}(\tilde{\Pi}_A^k\otimes I_B))$. We have
$D(\rho_{AB})\geq\sum_kD_H((I_A\otimes \Psi_B)\rho_{AB},(\tilde{\Pi}_A^k\otimes I_B)\rho_{AB}(\tilde{\Pi}_A^k\otimes I_B))\geq D((I_A\otimes \Psi_B)\rho_{AB})$.

3) For pure states, $D(\rho_{AB})$ is directly related to quantum entanglement. As $D(\rho_{AB})$ is not changed under the local unitary operations, we consider
the pure bipartite states $|\psi\rangle_{AB}$ in the form of Schmidt decomposition, $|\psi\rangle_{AB}=\sum_{i=0}^{r-1}\lambda_i|ii\rangle_{AB}$,
with $\lambda_i$ the Schmidt coefficients and $r=\min\{d_A,d_B\}$ the Schmidt rank. We have
\begin{eqnarray}\label{de}
D(\rho_{AB})&&=2\left(1-\sqrt{\max_{\{\Pi_A^k\}}\sum_k\Big|\sum_{i,j=0}^{r-1}\lambda_i\lambda_j|ii\rangle\langle jj|(|k\rangle\langle k|\otimes I_B)\Big|^2}\right)\nonumber\\
&&=2\left(1-\sqrt{\max_{\{\Pi_A^k\}}\sum_k\Big|\langle k|\sum_{i=0}^{r-1}\lambda_i^2|i\rangle_A\langle i|k\rangle\Big|^2}\right)\nonumber\\
&&\geq2\left(1-\sqrt{\sum_k^{r-1}\lambda_k^4}\right)=2\left(1-\sqrt{\mathrm{Tr}\rho_A^2}\right),
\end{eqnarray}
where $\rho_A$ is the reduced density matrix of $|\psi\rangle_{AB}$. It is obvious that $D(\rho_{AB})\geq2-\sqrt{4-2C^2(|\psi\rangle_{AB})}$, where $C(|\psi\rangle_{AB})=\sqrt{2(1-\mathrm{Tr}\rho_A^2)}$ is the concurrence
of $|\psi\rangle_{AB}$.

4)  The maximum of $D(\rho_{AB})$ is $2\left(1-\frac{1}{\sqrt{r}}\right)$ for $d_A\otimes d_B$-dimensional pure states $\rho_{AB}=|\psi\rangle_{AB}\langle\psi|$.
From (\ref{de}), for any pure state $|\psi\rangle_{AB}$ one has $D_{S}(\rho_{AB})=2\left(1-\sqrt{\sum_k\Big|\langle k|\sum_{i=0}^{r-1}\lambda_i^2|i\rangle_A\langle i|k\rangle\Big|^2}\right)\leq2\left(1-\sqrt{\sum_k\Big|\sum_{i=0}^{r-1}\lambda_i\langle i|k\rangle_A\langle k|i\rangle\Big|^4}\right)\leq2\left(1-\frac{1}{\sqrt{r}}\right)$,
where the first inequality is based on the convexity and the equality is saturated if all $\lambda_i$ are the same, i.e., $\lambda_i=\frac{1}{\sqrt{r}}$.
In other words, it is saturated by the maximally entangled states.

With the above properties of the quantum correlation measure $D(\rho_{AB})$, we claim the following theorem.

{\bf [Theorem 2]}.  $D(\rho_{AB})$ vanishes if and only if $\rho_{AB}$ is a classical-quantum correlated state.

{\bf [Proof]}. For a classical-quantum correlated state $\rho_{AB}=\sum_k\lambda_k|k\rangle\langle k|\otimes \rho_k$, one can always find
such $\Pi_A^k$ such that $\mathrm{Tr}\left(\sqrt{\rho_{AB}}-\sqrt{(\Pi_A^k\otimes I_B)\rho_{AB}(\Pi_A^k\otimes I_B)}\right)^2=0$ for all $k$.
On the contrary, given an arbitrary $\rho_{AB}$, if $\mathrm{Tr}\left(\sqrt{\rho_{AB}}-\sqrt{(\Pi_A^k\otimes I_B)\rho_{AB}(\Pi_A^k\otimes I_B)}\right)^2=0$ for some particular $k_1$ , we can write
$\rho_{AB}=\lambda_{k_1}|k_1\rangle\langle k_1|\otimes \rho_1+\rho_{k_1\perp}$ for some $\rho_{k_1\perp}$ and $\lambda_{k_1}\geq0$. Similarly, if
$\mathrm{Tr}\left(\sqrt{\rho_{AB}}-\sqrt{(\Pi_A^{k_j}\otimes I_B)\rho_{AB}(\Pi_A^{k_j}\otimes I_B)}\right)^2=0$ for all $k_j$ such that $\sum_j|k_j\rangle\langle k_j|=I_A$, we have
$\rho_{AB}=\sum_j\lambda_{k_j}|k_j\rangle\langle k_j|\otimes \rho_j$, which is obviously a classical-quantum correlated state.\quad $\Box$

As a detailed example, we now compute the quantum correlation $D(\rho_{AB})$ for arbitrary qubit-qudit states. For bipartite $2\otimes d$ states $\rho_{AB}$,
Eq. (\ref{defone}) can be rewritten as
\begin{eqnarray}\label{def}
D(\rho_{AB})&&=\mathrm{min}_{\{\Pi_A^k\}}\sum_kD_H(\rho_{AB},(\Pi_A^k\otimes I_B)\rho_{AB}(\Pi_A^k\otimes I_B))\nonumber\\
&&=2\left(1-\mathrm{max}_{\{\Pi_A^k\}}\sum_k\mathrm{Tr}\sqrt{\rho_{AB}(\Pi_A^k\otimes I_B)\rho_{AB}(\Pi_A^k\otimes I_B)}\right)\nonumber\\
&&=2\left(1-\mathrm{max}_{\{\Pi_A^k\}}\sum_k\mathrm{Tr}\sqrt{\rho_{AB}}(\Pi_A^k\otimes I_B)\sqrt{\rho_{AB}}(\Pi_A^k\otimes I_B)\right).
\end{eqnarray}
The second equality is due to $\Pi_A h(\rho)\Pi_A=h(\Pi_A\rho_{AB}\Pi_A)$ for any function $h$ of the qubit state in subsystem $A$  Ref. \cite{dtg}.

The projective measurement operator $\Pi_A^k$ can be expressed in Bloch representation,
\begin{eqnarray}\label{2d1}
\Pi_A^k=\frac{1}{2}(I_2+\vec{r}_k \cdot \vec{\sigma}),
\end{eqnarray}
where ${I}_2$ is the $2\times 2$ identity matrix, $\vec{r}_k\cdot \vec{\sigma}=\sum_{i=1}^3 {r}_k^i {\sigma_i}$ with $\sum_{i=1}^3 (r_{k}^i)^2=1$, $\sigma_i$ are Pauli matrices.
Substitute Eq. (\ref{2d1}) into Eq. (\ref{def}), one arrives at
\begin{eqnarray}\label{2d2}
D(\rho_{AB})=1-\mathrm{max}_{\{\vec{r_k}\}}\sum_{ij}r_{k_i}T_{ij}r_{k_j}=1-\lambda_{max},
\end{eqnarray}
where $\lambda_{max}$ is the maximal eigenvalue of the matrix $T$ with entries $T_{ij}=\mathrm{Tr}\sqrt{\rho_{AB}}(\sigma_i\otimes I_B)\sqrt{\rho_{AB}}(\sigma_j\otimes I_B)$.
Therefore, $D(\rho_{AB})$ can be analytically solved for qubit-qudit states, which is different from other non-classical correlations like quantum discord which has no analytical
formula even for two-qubit states \cite{ka}.
Interestingly, for this qubit-qudit case, the quantum correlation Eq. (\ref{2d2}) happens to be the local quantum uncertainty as the minimum skew
information achievable on a single local measurement introduced in \cite{dtg}.

\noindent{\bf Converting coherence to quantum correlations}~ With respect to the quantum correlation $D(\rho_{AB})$ given in (\ref{defone}),
the symmetric version of quantum correlation based on Hellinger distance for bipartite states $\rho_{AB}$ can be defined by
\begin{eqnarray}\label{def3}
D_S(\rho_{AB})=\mathrm{min}_{\{\Pi_A^k,\Pi_B^j\}}\sum_{k,j}D_H\left(\rho_{AB},(\Pi_A^k\otimes \Pi_B^j)\rho_{AB}(\Pi_A^k\otimes \Pi_B^j)\right),
\end{eqnarray}
where $D_H(\rho_{AB},(\Pi_A^k\otimes \Pi_B^j)\rho_{AB}(\Pi_A^k\otimes \Pi_B^j))=\mathrm{Tr}\left(\sqrt{\rho_{AB}}-\sqrt{(\Pi_A^k\otimes \Pi_B^j)\rho_{AB}(\Pi_A^k\otimes \Pi_B^j)}\right)^2$,
$\Pi_A^k=|k\rangle_A\langle k|$ and $\Pi_B^j=|j\rangle_B\langle j|$ denote the projective measurements on systems $A$ and $B$, respectively.

Let $\Lambda$ be an incoherent operation on a bipartite product state
$\rho_A\otimes \rho_B$. From Eq. (\ref{def3}), one has $D_S(\Lambda[\rho_A\otimes \rho_B])\leq C_H(\Lambda[\rho_A\otimes \rho_B])$.
Based on the monotonicity of the coherence, we arrives at
$C_H(\Lambda[\rho_A\otimes \rho_B])\leq C_H(\rho_A\otimes \rho_B)
=2\left(1-[1-\frac{1}{2}C_H(\rho_{A})][1-\frac{1}{2}C_H(\rho_{B})]\right)$.
Therefore, we have
\begin{eqnarray}\label{th4}
D_S(\Lambda[\rho_A\otimes \rho_B])\leq 2\left(1-[1-\frac{1}{2}C_H(\rho_{A})][1-\frac{1}{2}C_H(\rho_{B})]\right).
\end{eqnarray}

Eq. (\ref{th4}) characterizes the transformation of the local coherence to global quantum correlation under incoherent operations.
If one of the reduced state $\rho_A~(\rho_B)$ is incoherent, Eq. (\ref{th4}) becomes $D_S(\Lambda[\rho_A\otimes \rho_B])\leq C_H(\rho_B)~(C_H(\rho_A))$, which is similar to the main result in Ref. \cite{jbdv}.
Eq.(\ref{th4}) shows that the coherence in the initial state of $\rho_A\otimes \rho_B$ can be converted to the non-classical correlation
by a suitable incoherent operation $\Lambda=\sum_{ij}|i,i\oplus j\rangle\langle i,j|$.
The state $\rho_A\otimes \rho_B$ can be converted to a non-classically correlated state via incoherent operations if and only if at least one of $\rho_{A}$ and $\rho_{B}$ is not coherent.

As an example, consider $\rho_A=\rho_B=\frac{1}{2}(|0\rangle+|1\rangle)(\langle0|+\langle1|)$. Then
$C_H(\rho_A\otimes\rho_B)=2$. Take $\Lambda=I_2\oplus i\sigma_y$ to be the incoherent operation, where $\sigma_y$ is the Pauli matrix.
The maximal quantum correlations that can generated under $\Lambda$ is
$D_S(\Lambda[\rho_A\otimes\rho_B])=2(2-\sqrt{2})\leq 2$.
Eq. (\ref{th4}) can be readily extended to multipartite systems.

Next, we show that the upper bound in Eq. (\ref{th4}) is attainable. Let $\rho_B=|k\rangle\langle k|$. The initial state can be written as $\rho_0=\rho_A\otimes |k\rangle\langle k|$.
Under the incoherent operation $\Lambda=\{U_I=\sum_{ij}|i,i\oplus j\rangle\langle i,j|\}$, the state becomes $\rho_f=U_I\rho_0U_I^{\dagger}$. Consider the eigenstate decomposition of $\rho_A=\sum_i\lambda_i|\psi_i\rangle_A\langle\psi_i|$, with $|\psi_i\rangle=\sum_ja_j^i|j\rangle$ the eignstates in basis $\{|j\rangle\}$. $\rho_f$ can be written as
\begin{eqnarray*}
\rho_f&&=\sum_i\lambda_iU_I|\psi_i\rangle_A|k\rangle_B\langle\psi_i|_A\langle k|_BU_I^{\dagger}\\
&&=\sum_i\lambda_i\left(\sum_ja_j^i|jj\oplus k\rangle\right)\left(\sum_j\langle jj\oplus k|{a_j^i}^\ast\right).
\end{eqnarray*}
The quantum coherence of $\rho_A$ within the basis $\{|j\rangle\}$ is given by
\begin{eqnarray}\label{pfth34}
C_H(\rho_A)&&=2\left(1-\sqrt{\sum_j\left(\sum_i\sqrt{\lambda_i}|\langle j|\psi_i\rangle|^2\right)^2}\right)\nonumber\\
&&=2\left(1-\sqrt{\sum_j\left(\sum_i\sqrt{\lambda_i}|a_j^i|^2\right)^2}\right).
\end{eqnarray}
According to the definition of quantum correlation $D_S$, one has
\begin{eqnarray*}\label{pfth35}
(1-\frac{1}{2}D_S(\rho_f))^2&&=\max_{\{|kk'\rangle\}}\sum_{kk'}\left(\sum_i\sqrt{\lambda_i}\left(\sum_ja_j^i\langle kk'|jj\oplus k\rangle\right)\left(\sum_j\langle jj\oplus k|kk'\rangle{a_j^i}^\ast\right)\right)^2\nonumber\\
&&=\max_{\{|kk'\rangle\}}\sum_{kk'}\left(\sum_i\sqrt{\lambda_i}\langle kk'|P_kD_i\otimes I|\phi\rangle\langle\phi|P_kD_i^\ast\otimes I|kk'\rangle\right)^2\nonumber\\
&&=\max_{U,V}\sum_j\left(\sum_i\sqrt{\lambda_i}|\langle j|U^\dagger P_kD_iP_kV^\ast|j\rangle|^2\right)^2\nonumber\\
&&=\max_{U,V}\sum_j\left(\sum_i\sqrt{\lambda_i}\Big|\sum_k[U^\dagger]_{jk}a_k^i[V^\ast]_{kj}\Big|^2\right)^2,
\end{eqnarray*}
where $|\phi\rangle=\sum_j|jj\rangle,~D_i=\mathrm{diag}(a_0,a_1,\cdots)$, $P_k=\sum_j|k\oplus j\rangle\langle j|$, $U$ and $V$ are some unitary operators, and
we have taken into account that $\sum_ja_j^i |jj\oplus k\rangle=(P_kD_i\otimes I)|\phi\rangle$. By utilizing the Cauchy-Schwarz inequality in Eq. (\ref{pfth35}), one finds
\begin{eqnarray}\label{pfth36}
D_S(\rho_f)&&\geq2\left(1-\sqrt{\max_{U}\sum_j\left(\sum_i\sqrt{\lambda_i}\sum_k\big|[U^\dagger]\big|^2_{jk}\big|a_k^i\big|^2\right)^2}\right)\nonumber\\
&&\geq2\left(1-\sqrt{\max_{U}\sum_{jk}\big|[U^\dagger]\big|^2_{jk} \left(\sum_i\sqrt{\lambda_i}\big|a_k^i\big|^2\right)^2}\right)\nonumber\\
&&=2\left(1-\sqrt{\sum_j\left(\sum_i\sqrt{\lambda_i}\big|a_j^i\big|^2\right)^2}\right),
\end{eqnarray}
where the second inequality comes from the convexity and the extreme value is achieved when we select the optimal basis $\{|kk'\rangle\}=\{|jj\rangle\}$. Comparing Eq. (\ref{pfth34})
with Eq. (\ref{pfth36}), we have
$D_S(\rho_f)\geq C_H(\rho_A)$.
However, based on Eq. (\ref{th4}), we have $D_S(\rho_f)\leq C_H(\rho_A)$ for
$\rho_B=|k\rangle\langle k|$ and $U_I$.This means in this case $D_S(\rho_f)=C_H(\rho_A)$, which completes the proof.

\noindent{\bf Conclusion}~
In summary, we have presented a strongly monotonic coherence measure in terms of the Hellinger distance. The analytic expression is explicitly presented.
The relation between this coherence measure and the fidelity, as well as the connection with the geometric measure of quantum coherence have been obtained.
Employing this coherence measure we have shown a polygamy relation that dominates the coherence distribution among multipartite systems.
In addition, we have introduced a measure of quantum correlation based on Hellinger distance.
The analytical formula of quantum correlation for arbitrary qubit-qudit states has been derived.
Moreover, the trade-off relation between the quantum coherence and the quantum correlation has also been established, showing that
the local coherence can be converted to global quantum correlations under incoherent operations.

\bigskip
\noindent{\bf Acknowledgments}\, This work is supported by the NSF of China under Grant No. 11675113.

\end{document}